\documentclass[%
 reprint,
 amsmath,amssymb,
 aps,
]{revtex4-2}

\usepackage{graphicx}
\usepackage{dcolumn}
\usepackage{bm}
\usepackage{tabularray}
\usepackage{gensymb}
\usepackage{hyperref}

\begin{document}

\preprint{APS/123-QED}

\title{Improved ion bunch quality of conical target irradiated by ultra-intense and ultra-short laser}

\author{Wan-Qing Su$^{1,2,4}$}
\author{Xi-Guang Cao$^{2,4,5}$}\email[Corresponding author.]{Email:caoxg@sari.ac.cn}
\author{Chun-Wang Ma$^{1,3}$}\email[Corresponding author.]{Email:machunwang@126.com}
\author{Guo-Qiang Zhang$^{2,4,5}$}
\author{Hui-Ling Wei$^{1}$}
\author{Yu-Ting Wang$^{1}$}
\author{Chun-Yuan Qiao$^{1}$}
\author{Jie Pu$^{1}$}
\author{Kai-Xuan Cheng$^{1}$}
\author{Ya-Fei Guo$^{1}$}
\affiliation{$^{1}$College of Physics, Henan Normal University, Xinxiang 453007, China\\
$^{2}$Shanghai Advanced Research Institute, Chinese Academy of Sciences, Shanghai 201210, China \\
$^{3}$Institute of Nuclear Science and Technology, Henan Academy of Sciences, Zhengzhou 450015, China \\
$^{4}$Shanghai Institute of Applied Physics, Chinese Academy of Sciences, Shanghai 201800, China\\
$^{5}$University of Chinese Academy of Sciences, Beijing 100049, China
}

\begin{abstract}
We conduct particle-in-cell simulations to estimate the effects of circularly and linearly polarized SEL 100 PW lasers on flat Th targets with thicknesses of 50 nm, 100 nm and 250 nm, as well as easy to manufacture conical Th targets with angularity either on the left or right. As the thickness of the three types of targets increases and under the same polarized laser, the average energy, maximum energy and energy conversion efficiency of Th ions decrease as it is well-known, and except for the circularly polarized laser hit on the conical target with angularity on the left, the Th ion beam emittance also decreases, while its beam intensity increases conversely. The linearly polarized laser, compared to the circularly polarized laser with the same laser intensity, exhibits higher beam intensity, beam emittance and energy conversion efficiency for the same type and thickness of Th target. The conical Th target with angularity on the left and intermediate thickness, compared to the flat target and conical target with angularity on the right of the same thickness, possesses both higher ion average energy up to 7 GeV and virtually the same beam intensity up to 0.8 MA under the linearly polarized laser. The results lead us to an easier way of controlling laser-accelerated high-quality heavy ion beam by switching to an optimal laser-target configuration scheme, which may enable the synthesis of superheavy nuclei in a high-temperature and high-density extreme plasma environment in astronuclear physics.
\begin{description}
\item[Keywords]
particle-in-cell, high-quality ion bunch, patawatt laser, superheavy nuclei, laser plasma accelerator
\end{description}
\end{abstract}

\maketitle

\section{\label{sec:introduction}INTRODUCTION}

With the development of laser technology, ultra-intense and ultra-short lasers with high energy density and short pulse duration have been increasingly applied in various aspects of physics research, which include fast ignition in laser-driven fusion \cite{tabak1994ignition}, diagnostics of high energy density physics \cite{borghesi2002laser, shen2020recent, lonardoni2022first}, compact electron and ion accelerators \cite{eliezer2012relativistic, wanqing2024multi}, laboratory-based radiobiological \cite{yogo2009application, bin2022anew, wang2022167196}, space research \cite{daido2012review}, material processing \cite{boody1996laser, yang2019fabrication} and high resolution radiography \cite{theobald2006hot, courtois2011high, do2022high}. The significant features of laser-driven ion beam are collimated and highly laminar with extremely small effective source size (around 10 $\mu m$) and picosecond-level even femtosecond-level ultra-short pulses \cite{limpouch2013laser}. The Station of Extreme Light (SEL), which is known as one of the end Stations of the Hard X-ray Free Electron Laser System (SHINE) being under construction in Shanghai, China, will provide ultra-intense and ultra-short laser up to 100 PW \cite{wang2019high, shao2020broad, hu2021numerical}. Laser particle accelerators have been proposed as crucial alternatives to traditional accelerators, offering potential advantages in terms of compactness, simplicity and cost \cite{bulanov2002oncological, he2022topology}.

Compound nuclei in excited states have been produced in experiments through fusion reactions between fast heavy ions in laser-accelerated plasmas and given atoms in adjacent activated samples \cite{mckenna2004characterization}. The $^{10}B$($\alpha$,n)$^{13}N$ reaction in an inertial confinement fusion implosion of the National Ignition Facility (NIF) has been measured for the first time, which is a preliminary step in the development of radiochemical mixed diagnosis \cite{lonardoni2022first}. Due to the unprecedented ion density of laser-accelerated ion beams, which is about 14 orders of magnitude higher than those from traditional accelerators, new fission-fusion reactions have the potential to produce extremely exotic isotopes near the waiting point region of the r-process at N = 126 \cite{thirolf2011laser}. Laser-accelerated ion beams are beneficial for observing multiple rapid reactions in laboratory astrophysics and measuring exotic cross-sections \cite{burggraf2022lasers, wei2021testing}. With increasing energy of laser-accelerated ion beams, optimization of beam quality and improvement in stability, it is anticipated that the superheavy island, which refers to the production of new superheavy nuclei beyond element 118 (Oganesson), can be accessed \cite{OGANESSIAN1995823, ma2020new, xin2021properties}. Recently the NIF has successfully achieved controlled nuclear fusion ignition with net energy gain, continuously surpassing previous levels of efficiency and precision, and a capsule target is subjected to the focus of 192 high-energy laser beams, causing the target to implode and achieve deuterium–tritium fusion \cite{zylstra2022burning}. The shape of the target plays a contributing role in laser acceleration. Near-conical solid targets of different shapes, such as double-cone target \cite{zhang2020double}, cone-in-shell target \cite{lei2006optimum}, nanobrush target \cite{cao2010enhanced, zhao2016monte, wang2023detection}, spherical thin-shell \cite{xu2012production, xu2014high, dahiya2015influence}, compound cone target \cite{king2009studies, yabuuchi2013impact, yasen2019enhancement}, bulged flat target \cite{zhang2012hundreds}, subwavelength multihole structured target \cite{nishiura2020control}, nanospheres \cite{seiffert2022strong}, wire target \cite{zhang2020generation}, etc., have been widely used in simulations and experiments to generate higher quality electron and ion beams, which specifically include the energy conversion efficiencies from laser to the particles, the beam collimation, the particle energy control, etc.

Further exploration is needed to product high-quality heavy ion beams by laser irradiating metal targets. Among naturally available heavy elements, Th and U possess the highest charge numbers, with Th being more available than U. Transfer reactions between heavy nuclei are a promising pathway for synthesizing superheavy nuclei, requiring both the shell and target nuclei to have high charge and mass numbers, and the selection of Th for acceleration can be also used to study the nuclear clocks \cite{zhang2024frequency}.
In contrast to commonly used flat targets (FT) \cite{domanski2018ultra}, we will construct conical Th targets with angularity pointing to the left and the right to obtain higher quality Th ion beams. The content of this paper is as follows: Sec.~\ref{sec:model} introduces the physical model and parameter settings used in particle-in-cell (PIC) simulations; Sec.~\ref{sec:result} presents the simulation results of Th ions, including the spatial and angular distributions, energy and charge spectra and spatial electric field distributions; Sec.~\ref{sec:ionbeam} discusses the ion beam parameters of different shapes of Th targets under different polarized lasers; and Sec.~\ref{sec:conclusion} is the conclusion.

\section{\label{sec:model}MODELS AND SIMULATIONS}

The laser-driven ion acceleration mechanisms mainly include target normal sheath acceleration (TNSA) \cite{snavely2000intense, hatchett2000electron, wilks2001energetic}, breakout afterburner (BOA) \cite{yin2006gev} and radiation pressure acceleration (RPA) \cite{pegoraro2007photon}. RPA can further be divided into laser-piston \cite{esirkepov2004highly}, hole-boring \cite{robinson2009relativistically, robinson2009hole}, light-sail \cite{simmons1993marx, macchi2009light} and phase-stable acceleration \cite{yan2008generating}. Based on the theoretical foundation of the aforementioned mechanisms, PIC simulations can reflect the real interaction process between ultra-intense and ultra-short laser and plasma.

Smilei is a collaborative, open-source, user-friendly PIC code for plasma simulation, which could be applied to a wide range of physics studies: from relativistic laser-plasma interaction to astrophysical plasmas \cite{derouillat2018smilei}. Particles satisfying Vlasov's equation are ionized in the electromagnetic field, and form a self-consistent dynamical system gradually \cite{perelomov1966ionization, perelomov1967ionization, ammosov1986tunnel}. The main theoretical formulas are,
\begin{equation}
\left [ \partial_t + \frac{\mathbf{p}}{m_{s}\gamma} \cdot \bigtriangledown + q_{s} \left ( \mathbf{E} + \mathbf{v}\times \mathbf{B} \right ) \cdot \bigtriangledown_{p} \right ]f_{s} = 0,
\end{equation}
\begin{equation}
\bigtriangledown \cdot \mathbf{B} = 0,
\end{equation}
\begin{equation}
\bigtriangledown \cdot \mathbf{E} = \rho / \epsilon_{0},
\end{equation}
\begin{equation}
\bigtriangledown \times \mathbf{B} = \mu_{0} \mathbf{J} + \mu_{0} \epsilon_{0} \partial_t \mathbf{E},
\end{equation}
\begin{equation}
\bigtriangledown \times \mathbf{E} = - \partial_t \mathbf{B},
\end{equation}
where $s$ denotes a given species consisting of particles with charge $q_{s}$ and mass $m_{s}$,
$\mathbf{x}$ and $\mathbf{p}$ respectively denote the position and momentum of a phase-space element, $f_{s}\left ( \mathbf{x}, \mathbf{p}, t\right )$ is distribution functions of the plasma; $\gamma = \sqrt{1 + \mathbf{p}^{2} / \left ( m_{s}c\right )^{2}}$ is the relativistic Lorentz factor, $c$ is the speed of light in vacuum; and $\epsilon_{0}$, $\mu_{0}$, $\rho$ and $\mathbf{J}$ are the vacuum permittivity and permeability, and charge and current densities, respectively.

\begin{figure}[htbp]
\includegraphics{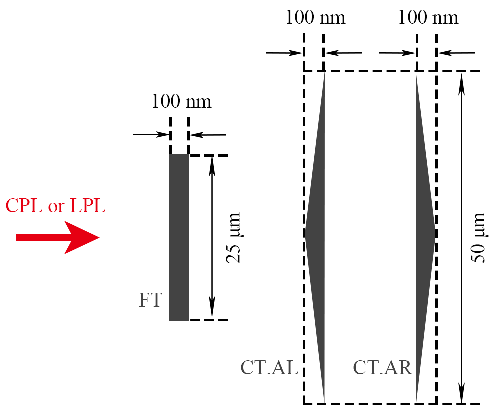}
\caption{(Color online) Schematic of SEL 100 PW circularly polarized laser (CPL) or linearly polarized laser (LPL) hitting on the flat target (FT), the conical target of angularity on the left (CT.AL) or the conical target of angularity on the right (CT.AR) with thicknesses of 100 nm.}
\label{target}
\end{figure}

According to the SEL 100 PW \cite{wang2019high, shao2020broad, hu2021numerical}, the peak intensity of the laser is set to $10^{23}$ W/cm$^2$, the full-width-at-half-maximum (FWHM) is 15 fs, focal spot size is 5 $\mu$m, and wavelength is $\lambda$ = 1 $\mu m$. Gaussian laser polarization can be either circular or linear, corresponding to the normalized vector potential of either $a$ = 190 or 269, respectively. The spatial size of the 2D3V (two-dimensional and three-velocity) PIC simulation is 40 $\lambda$ $\times$ 100 $\lambda$, with a spatial step of 12.5 nm and a time step of 22 as. The Th target is located at $X$ = 10 $\lambda$. Different target thicknesses, 50 nm, 100 nm, which is shown in Fig.~\ref{target}, and 250 nm, correspond to different shapes, which are flat target (FT), conical target with angularity on the left (CT.AL) and conical target with angularity on the right (CT.AR) with the same area $S$ = 2.5 $\lambda^2$. Each grid contains 49 Th ions and 49 electrons. The Th ion density is $n = 3 \times 10^{28} /m^3$ with every ion charge of 1+, and the electron density is set to be the same as the adjacent density $n_e = 1.1 \times 10^{27} /m^3$. Laser irradiates at the center position of three different shaped targets, and particles move and ionize gradually in the electromagnetic field based on the above theoretical model through iterative calculations. 

\section{\label{sec:result}RESULTS}

\subsection{\label{sec:xy}Ion spatial and angular distributions}

\begin{figure}[htbp]
\includegraphics{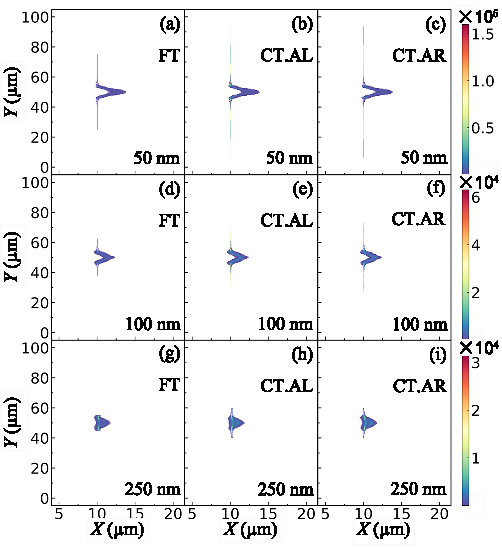}
\caption{(Color online) Ion spatial distributions simulated by Smilei for the circularly polarized SEL 100 PW laser on Th target at 100 fs. (a)-(c) The flat target (FT), the conical target of angularity on the left (CT.AL) and the conical target of angularity on the right (CT.AR) with thicknesses of 50 nm, respectively. (d)-(f) FT, CT.AL and CT.AR with thicknesses of 100 nm, respectively. (g)-(i) FT, CT.AL and CT.AR with thicknesses of 250 nm, respectively. See Sec. \ref{sec:ionbeam} for beam parameters.}
\label{cir-xy}
\end{figure}

\begin{figure}[htbp]
\includegraphics{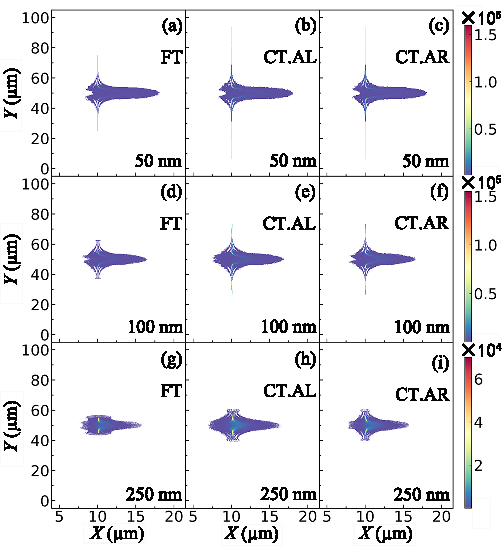}
\caption{(Color online) Ion spatial distributions simulated by Smilei for the linearly polarized SEL 100 PW laser on Th target at 100 fs. (a)-(c) The flat target (FT), the conical target of angularity on the left (CT.AL) and the conical target of angularity on the right (CT.AR) with thicknesses of 50 nm, respectively. (d)-(f) FT, CT.AL and CT.AR with thicknesses of 100 nm, respectively. (g)-(i) FT, CT.AL and CT.AR with thicknesses of 250 nm, respectively. See Sec. \ref{sec:ionbeam} for beam parameters.}
\label{lin-xy}
\end{figure}

When the polarized SEL 100 PW laser interacts with corresponding target of different thicknesses, the spatial distribution of Th ions in the conical target (CT), compared to the FT, is shown in Fig.~\ref{cir-xy} and Fig.~\ref{lin-xy} at 100 fs. As shown in Fig.~\ref{cir-xy}(a)-(c), under the conditions of the circularly polarized laser (CPL) and the thinnest target, the spatial distribution of Th ions in the three different shaped targets is akin. However, the CT.AL has a higher ion density at the target center compared to the CT.AR, and the FT has the lowest ion density at the target center. The trend of Th ion density of the intermediate thickness target, as shown in Fig.~\ref{cir-xy}(d)-(f), is similar to that of the thinnest target at the target center. Due to the increased target thickness, the ions accelerated by the laser are slightly pushed away and closer to the initial position, while a clear image of the RPA acceleration mechanism could still be observed. In the thickest and thinnest target, the density of Th ions at the center of the CT is greater than those of the FT. Distinct differences of the three different shaped targets can be observed in Fig.~\ref{lin-xy}, which shows the spatial distribution of Th ions on the edges of the targets under the linearly polarized laser (LPL) at 100 fs, and the CT.AL experiences more significant acceleration at its edge. Although the thick target is not easily pushed, the thickest CT.AL exhibits the most effective overall acceleration compared to the thickest CT.AR and FT. In the thinnest target, most of the Th ions accelerated at the center move forward, and a small portion disperses around the target center, and the ion spatial distributions of the three shaped targets are also akin. In the intermediate thickness targets, the spatial distributions of Th ions accelerated in the CT.AR and FT are almost the same, and the number of ions accelerated in the CT.AL is lowest, and there are most of ions located around the center of CT.AL. In the thickest target, Fig.~\ref{lin-xy}(g)-(i) clearly show that the CT.AR has most of Th ions being accelerated forward at the center, and the ions accelerated in the CT.AL and FT are mostly located around the target center. The LPL causes more significant expansion and acceleration of the Th target than those in the CPL pattern.

\begin{figure*}[htbp]
\includegraphics{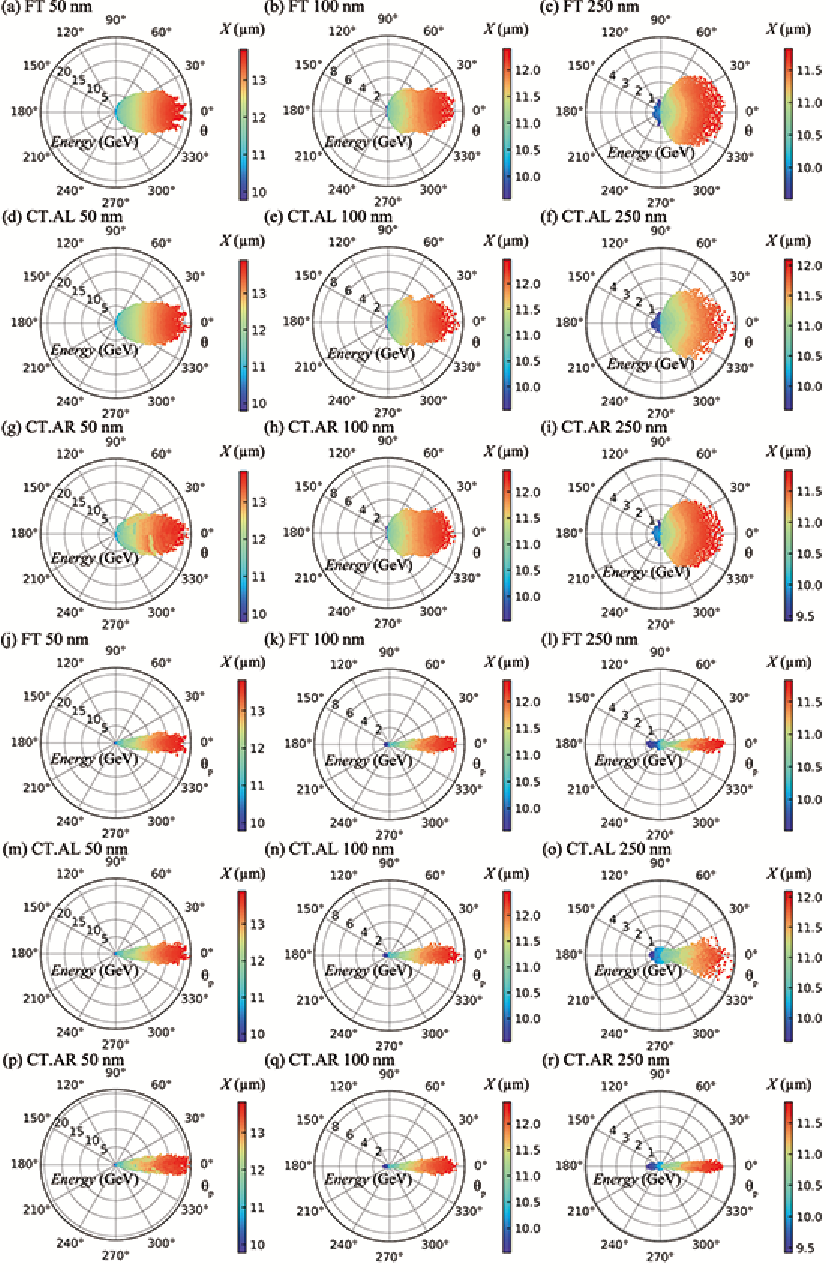}
\caption{(Color online) The distributions of spatial angle ($\theta$) and momentum angle ($\theta_p$) corresponding to energy ($Energy$) and space ($X$) simulated by Smilei for the circularly polarized SEL 100 PW laser on Th target at 100 fs. (a)-(c) and (j)-(l) The foil targets (FTs) with thicknesses of 50 nm, 100 nm, and 250 nm, respectively. (d)-(f) and (m)-(o) The conical targets of angularity on the left (CT.ALs) with thicknesses of 50 nm, 100 nm, and 250 nm, respectively. (g)-(i) and (p)-(r) The conical targets of angularity on the right (CT.ARs) with thicknesses of 50 nm, 100 nm, and 250 nm, respectively.}
\label{Ang-cir}
\end{figure*}

\begin{figure*}[htbp]
\includegraphics{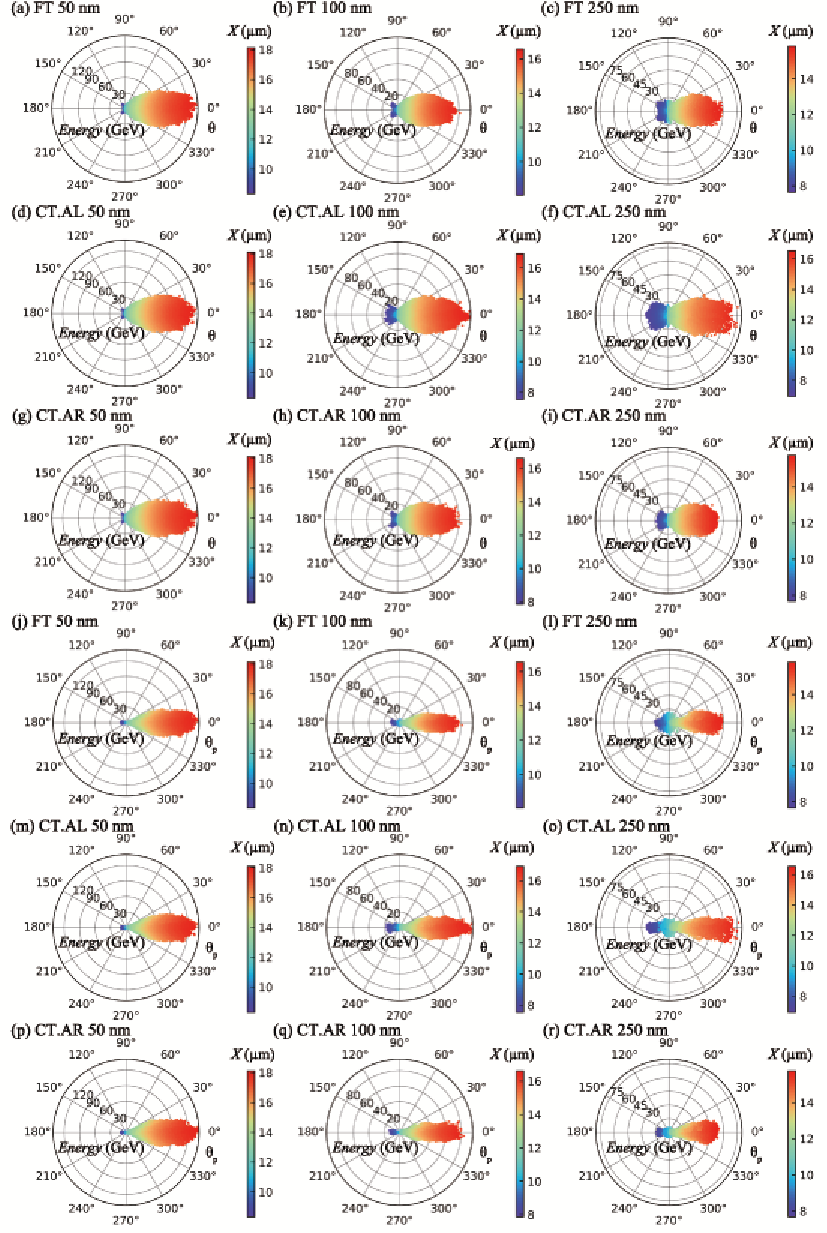}
\caption{(Color online) The distributions of spatial angle ($\theta$) and momentum angle ($\theta_p$) corresponding to energy ($Energy$) and space ($X$) simulated by Smilei for the linearly polarized SEL 100 PW laser on Th target at 100 fs. (a)-(c) and (j)-(l) The foil targets (FTs) with thicknesses of 50 nm, 100 nm, and 250 nm, respectively. (d)-(f) and (m)-(o) The conical targets of angularity on the left (CT.ALs) with thicknesses of 50 nm, 100 nm, and 250 nm, respectively. (g)-(i) and (p)-(r) The conical targets of angularity on the right (CT.ARs) with thicknesses of 50 nm, 100 nm, and 250 nm, respectively.}
\label{Ang-lin}
\end{figure*}

The spatial angle ($\theta = \arctan (y/x)$) and momentum angle ($\theta_p = \arctan (p_y/p_x)$) of Th ions, along with their relationship with energy and $X$ axis coordinates, are shown in the Fig.~\ref{Ang-cir} and Fig.~\ref{Ang-lin}, which can be observed that the ion energy is highest when the angles of spatial and momentum are near $0\degree$, and the Th ion conversely has lower energy when the spatial angle $\theta$ approaches $90\degree$ and $270\degree$. In the CPL pattern, there are more Th ions and higher energy for the spatial angle $\theta$ approaches $180\degree$ as target thickness increasing, and ions with momentum angle $\theta_p$ closer to $180\degree$ have higher energy compared to those on the upper and lower ends. Especially for the CT.AL, the angular divergence of Th ions increases as the thickness increases, while for the CT.AR and FT, the momentum angular divergence of ions decreases as the thickness increases. In the LPL pattern, the trend of angular distributions is quite similar to that in the CPL pattern, while the Th ion energy is notably lower near the spatial angle $\theta$ of $180\degree$ than those around $120\degree$ and $240\degree$ for the CT.ARs and FTs, and the ion energy consistently increases at the center and the upper and lower ends of CT.ALs. Conversely for the CT.AL, the momentum angular divergence of Th ions decreases significantly and the ion energy than that of the CT.AR and FT as the thickness increases. 

\subsection{\label{sec:Ekin}Ion energy and ionization}

\begin{figure*}[htbp]
\includegraphics{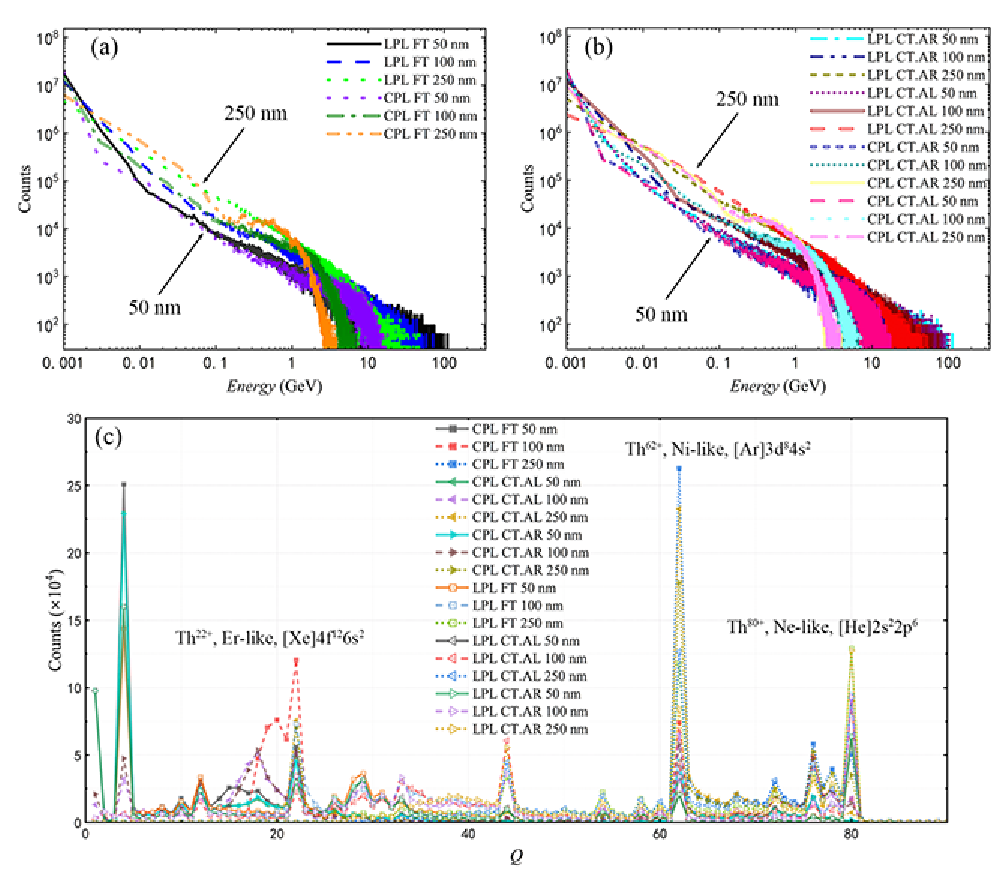}
\caption{(Color online) Ion energy and charge spectra simulated by Smilei for both circularly polarized and linearly polarized SEL-100 PW laser on Th target at 100 fs. (a) and (b) The Th ions energy spectra of FTs and CTs, respectively. (c) The Th ions charge spectra with peaks include Th$^{22+}$, Th$^{62+}$ and Th$^{80+}$, respectively.}
\label{Ekin&Q}
\end{figure*}

At 100 fs, the energy and charge spectra of all Th ions can be seen in Fig.~\ref{Ekin&Q}. As Fig.~\ref{Ekin&Q}(a) and (b) shows, the Th ions energy spectra of the three shaped targets with the same thickness almost overlap within the low-energy range of 0-100 MeV in the LPL and CPL pattern, but the thinner the target, the more ions there are in the low-energy region, and the faster the energy spectrum decreases. The Th ions accelerated by the CPL have a distinct energy peak compared to those accelerated by the LPL in the mid-energy range of 0.1-10.0 GeV, and the peak energy of energy spectrum decreases and the number of peak ions increases with increasing target thickness. The Th ions accelerated by the LPL have higher maximum energies compared to those accelerated by the CPL, and the thinner the target, the higher the maximum energy of the Th ions in the same polarized laser pattern. Overall, the energy spectrum of Th ions is narrower with an energy peak in the CPL pattern, and there is a wider energy spectrum in the LPL pattern, and the ion energy spectra are similar for three shaped targets with the same thickness in the same polarized laser pattern.

Fig.~\ref{Ekin&Q}(c) displays that the charge of Th ions is mostly concentrated on 62+, followed by 22+ in the CPL pattern, while the charge of Th ions is mostly concentrated on 80+, followed by 62+ in the LPL pattern. The ionization of Th ions in the CPL pattern is also relatively lower compared to that in the LPL pattern. For Th$^{22+}$, the target that produces it the most is the 100 nm FT in the CPL pattern with about $1.2\times10^5$, followed by the 250 nm CT.AL in the CPL pattern with approximately $8\times10^4$, and closely behind are the 100 nm FT in the LPL pattern and 250 nm CT.AR in the CPL pattern, both with about $7\times10^4$. For Th$^{62+}$, the target that produces it the most is the 250 nm FT in the CPL pattern with about $2.6\times10^5$, followed by the 250 nm CT.AL in the CPL pattern with about $2.3\times10^5$, and behind is the 250 nm CT.AR in the CPL pattern with about $1.8\times10^5$. For Th$^{80+}$, the target that produces it the most is the 250 nm FT and CT.AR in the LPL pattern with about $1.3\times10^5$, followed by the 100 nm FT, CT.AR and CT.AL and the 250 nm CT.AL in the LPL pattern, all with about $9\times10^4$. Other targets also produce small amounts of the above three ions. Additionally, the Th target with the thinnest thickness has the highest concentration of ions in the 4+ charge state, and the peak charge of Th ions charge spectrum increases as the target gets thicker.

\subsection{\label{sec:Ex}Space electric field}
 
\begin{figure}[htbp]
\includegraphics{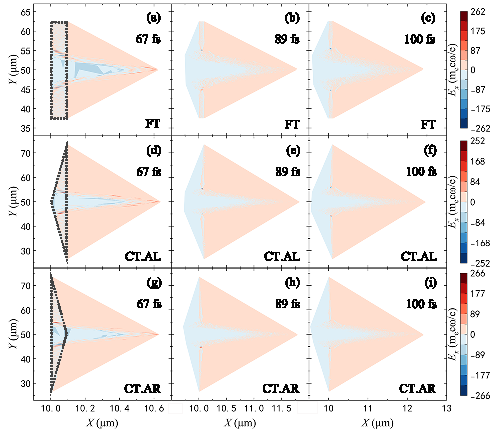}
\caption{(Color online) The spatial electromagnetic field distribution simulated by Smilei for the circularly polarized SEL 100 PW laser on 100 nm Th target at 67 fs, 89 fs and 100 fs, respectively. (a)-(c) The FT. (d)-(f) The CT.AL. (g)-(i) The CT.AR.}
\label{Excir}
\end{figure}

\begin{figure}[htbp]
\includegraphics{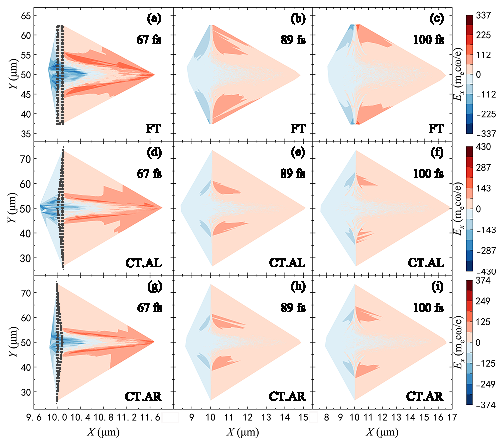}
\caption{(Color online) The spatial electromagnetic field distribution simulated by Smilei for the linearly polarized SEL 100 PW laser on 100 nm Th target at 67 fs, 89 fs and 100 fs, respectively. (a)-(c) The FT. (d)-(f) The CT.AL. (g)-(i) The CT.AR.}
\label{Exlin}
\end{figure}

By selecting the electric field distribution of the target with intermediate thickness, which can reflect the real-time evolution of ion motion corresponding to 67 fs, 89 fs and 100 fs, we can enhance the understanding of the spatial and angular distribution and the energy and charge of Th ions in different shaped targets.

In the CPL pattern (Fig.~\ref{Excir}), the CT.AL has a strongest electric field at the front, which is the end of the positive direction of the $X$ axis, and its electric field is distributed at the furthest front. The CT.AR has a similar electric field distribution to the FT, while there is a sharp corner behind the target at 89 fs and 100 fs. In the LPL pattern (Fig.~\ref{Exlin}), the CT.AL still has the strongest positive electric field, and its electric field is also distributed at the furthest front, while its negative electric field distribution is the widest. The electric field evolution of the CT.AR is similar to that of the FT as well, and the electric field distribution is weaker at the upper and lower ends of the CT than at the center surrounding.

Compared with the electric field in the CPL pattern, the electric field in the LPL pattern not only has a greater intensity, but the distribution of negative electric field towards the rear of target is wider, and the electric field on both sides of the $X$ axis has a more notable weakening as laser moves away from the target center. The acceleration mechanism of Th ions is clearly manifested as RPA in this simulation, where the target is integrally accelerated by the ultra-intense and ultra-short laser, and its contour closely resembles the shape of the laser profile with the considerable effect of the targets with different thicknesses, i.e., the CTs with different angularities. This mechanism is particularly remarkable for Th target in the CPL pattern \cite{macchi2005laser, zhang2007efficient, xu2007vacuum}, while the LPL is more effective in promoting Th ions, allowing them to obtain higher energy and ionization.

\section{\label{sec:ionbeam}DISCUSSION}

Regarding the prior simulation of Th ion beam, J. Doma\'nski et al. have obtained a multi-GeV heavy ion beam by using a laser with an intensity of $10^{23}$ W/cm$^2$ (E = 150 J) striking a FT \cite{domanski2018ultra}. Our simulation reveals that particle acceleration could be controlled to achieve a higher quality Th ion beam by adjusting the target shape and laser polarization, for instance, by using the CTs with different thicknesses and the LPL. In the selecting the beam region of $X \in$ [10.5, 40.0] $\mu m$ and $Y \in$ [45.0, 55.0] $\mu m$, we then amply discuss key parameters of the Th ion beam, including beam intensity, beam emittance, beam ion energy, beam ion charge and energy conversion efficiency. According to the spatial distribution and ionization of heavy ions, 2D space could be extended to 3D space. The corresponding ion beam intensity is obtained from Eq. (\ref{eq:L}),
\begin{equation}
L= \frac{N v \overline{Q} e}{l}\label{eq:L},
\end{equation}
where $l$ is the length of the chosen space in the $X$ direction; and $N$ is the number of ions, $v$ is the average velocity of ions, $\overline{Q}$ is the average charge number of ions in the chosen space, respectively; and $e$ is the unit electric charge.
The emittance of the ion beam in 2D space is described by Eq. (\ref{eq:D}),
\begin{equation}
\varepsilon = \sqrt{\left \langle y^{2}\right \rangle\left \langle \left ( \frac{p_{y}}{p_{x}}\right )^{2}\right \rangle - \left \langle y \tfrac{p_{y}}{p_{x}}\right \rangle^{2}}\label{eq:D}.
\end{equation}
The beam intensity and emittance of Th ions in the beam region are obtained from the above equations with the respective results of simulations at $100  ~\mathrm{fs}$. The conversion efficiency $\eta $ can be obtained by $E_{ions}$ / $E_{laser}$, where $E_{ions}$ is the energy of all Th ions and $E_{laser}$ $\approx$ 424 J is the energy of the above SEL 100 PW laser. The simulation results of those parameters are listed in Tab.~\ref{tab:ParaIonBeam}.

\begin{table*}
\centering
\caption{\label{tab:ParaIonBeam}Table of Th ion beams parameters simulated by Smilei for SEL 100PW laser, include laser polarization (P); target thickness (D); target shape (T) as the foil target (FT) , the conical target of angularity on the left (CT.AL) and the conical target of angularity on the right (CT.AR); beam intensity (L); beam emittance ($\varepsilon$); average beam ion energy (E$_{avg}$); maximum beam ion energy (E$_{max}$); average beam ion charge (Q$_{avg}$); and energy conversion efficiency ($\eta$).}
\begin{tblr}{
  cell{2}{1} = {r=9}{},
  cell{2}{2} = {r=3}{},
  cell{5}{2} = {r=3}{},
  cell{8}{2} = {r=3}{},
  cell{11}{1} = {r=9}{},
  cell{11}{2} = {r=3}{},
  cell{14}{2} = {r=3}{},
  cell{17}{2} = {r=3}{},
}
\hline
P & D (nm)  & T & L (MA) & $\varepsilon$ (mm$\times$mrad) & E$_{avg}$ (GeV) & E$_{max}$ (GeV) & Q$_{avg}$ (e) & $\eta$ ($\%$)       \\
\hline
CPL & 50  & FT & 0.289 & 0.105 & 3.65 & 20.54 & 71 & 8.765 \\
CPL & 50  & CT.AL & 0.287 & 0.107 & 3.66 & 21.75 & 71 & 8.778 \\
CPL & 50  & CT.AR & 0.290 & 0.103 & 3.64 & 20.99 & 71 & 8.764 \\
\hline
CPL & 100 & FT & 0.412 & 0.081 & 1.27 & 7.98 & 76 & 5.270 \\
CPL & 100 & CT.AL & 0.410 & 0.119 & 1.30 & 8.52 & 75 & 5.353 \\
CPL & 100 & CT.AR & 0.411 & 0.084  & 1.29 & 8.04 & 75 & 5.323  \\
\hline
CPL & 250 & FT & 0.429 & 0.045 & 0.77 & 4.23 & 75 & 4.084 \\
CPL & 250 & CT.AL & 0.407 & 0.146 & 0.73 & 5.06 & 70 & 3.958 \\
CPL & 250 & CT.AR & 0.427 & 0.044 & 0.78 & 4.20 & 75 & 4.084 \\
\hline
LPL & 50  & FT & 0.624 & 6.235 & 14.15 & 140.81 & 76 & 37.851 \\
LPL & 50  & CT.AL & 0.623 & 19.574 & 14.26 & 136.27 & 76 & 37.963 \\
LPL & 50  & CT.AR & 0.623 & 8.363 & 14.27 & 140.61 & 76 & 38.041 \\
\hline
LPL & 100 & FT & 0.875 & 2.094 & 5.51 & 81.19 & 77 & 25.545 \\
LPL & 100 & CT.AL & 0.844 & 9.988 & 6.68 & 94.31 & 76 & 29.248 \\
LPL & 100 & CT.AR & 0.858 & 2.352 & 5.91 & 83.22 & 77 & 26.182 \\
\hline
LPL & 250 & FT & 1.195 & 0.356 & 2.52 & 58.86 & 74 & 20.719 \\
LPL & 250 & CT.AL & 1.187 & 0.221 & 3.20 & 76.47 & 70 & 27.411 \\
LPL & 250 & CT.AR & 1.134 & 0.362 & 2.71 & 54.81 & 75 & 20.402 \\
\hline
\end{tblr}
\end{table*}

The Th ion beam intensities produced by mid-thickness and thickest targets are significantly higher than those generated by thinnest targets, and beam intensity increases as the target thickness increases, except for the mid-thickness CT.AL in the CPL pattern, which has a slightly higher beam intensity than the thickest target. This discrepancy could be attributed to the certain angle of the contact surface between the laser and the CT.AL, and its acceleration process is significantly different from those of the CT.AR and FT. Additionally, the difference in beam intensity of Th targets with different thicknesses in the CPL pattern is relatively smaller than that in the LPL pattern, and the ion beam intensity in the LPL pattern is higher than that in the CPL pattern. The beam emittance decreases as the target thickness increases, except for the CT.AL in the CPL pattern, whose emittance increases with increasing thickness. In the LPL pattern, the target is more expansive and results in higher beam emittance compared to the CPL pattern. Those results may be explained by the fact that the normalized vector potential of the LPL is higher.

Both the average energy and maximum energy of Th ions in the LPL pattern are higher than those in the CPL pattern with the same laser energy, and they decreases as the target thickness increases. Intriguingly, the Th ions of the CT have higher energy compared to those of the FT. For the CPL and the thinnest and mid-thickness targets, the Th ions of the CT.AL have the highest average and maximum energies. For the CPL and the thickest targets, the Th ions of the CT.AL have the lowest average energy and highest maximum energy. For the LPL and mid-thickness and thickest targets, the Th ions of the CT.AL have the highest average and maximum energies. For the LPL and the thinnest targets, the Th ions of the CT.AL have the lower average energy and lowest maximum energy. The differences between the average and maximum ion energies are likely due to laser polarization and the shape and thickness of target. The average charge of Th ions in different scenarios is either greater than or equal to 70+, indicating that the beam consists of highly charged ions. Remarkably, the energy conversion efficiency of the LPL pattern is approximately five times higher than that of the CPL pattern, and it increases as the target thickness decreases, while it is highest for mid-thickness CT.AL and both CPL and LPL because the CT.AL has the largest contact surface with the laser.

Taking into account the above parameters of the Th ion beam, the mid-thickness CT.AL simultaneously exhibits high beam intensity, energy and energy conversion efficiency either in the CPL or LPL pattern. However, the thinnest target can be chosen if a higher ion beam energy is desired, and the thickest target can be selected if a higher beam intensity is required. It is noteworthy that the CT.AR and FT are preferable compared to the CT.AL when using the CPL, and the CT.AL is preferable over the CT.AR and FT when using the LPL.

Although compared to the short target, the beam intensities of Th ions produced by these three shapes long targets are lower and the beam emittance is larger \cite{wanqing2024multi}, those targets are easier to manufacture and use selectively in experiments, and the findings apply to other metal ions as well \cite{domanski2018ultra}. Non-planar targets can further improve the quality of the ion beam. The left protrusion, the interaction plane not perpendicular to the laser direction, such as the CT.AL, could significantly increase the ion maximum energy and the energy conversion efficiency. The right protrusion, the interaction plane perpendicular to the laser direction, such as the CT.AR, is similar to the FT, and might also effectively increase the ion average energy. Based on the above regularities that Th ions with larger $X$ axis coordinates generally have higher energies in the beam and Th ion beam parameters vary corresponding to different polarized lasers and target thicknesses and shapes, it is prospective to discuss the following applications of Th ion beams in different energy ranges. The ion beam is mostly composed of low-energy ions, which could be used to simulate astrophysical phenomena like supernova explosions \cite{darke2000laser, travaglio2013nucleosynthesis}. Especially, medium-energy ions have an energy peak under CPL, and it also contains a larger number of medium-energy Th ions, in which ions of 0.1-1.0 GeV could be utilized in studying nuclear matter symmetry energies, nuclear reaction dynamics, multifragmentation, exotic nuclear structures and isotope production like $^{99}$Mo \cite{grismold2016large}, and ions of 1-10 GeV could be utilized in studying quark-gluon degrees of freedom and antimatter nuclei. Meanwhile, high-energy ions, which might be preferred for heavy-ion collisions with those ions accelerated by LPL being the optimal choice due to their high maximum energy, are the fewest in ion beam \cite{zhang2020progress, abdulhamid2024observation}. Due to the limited thickness of the targets under the RPA, the angularities of the three selected targets with different thicknesses are relatively large, and further exploration is needed for more optimal shape targets and laser incidence angles to qualitatively improve the parameters of the heavy ion beam and promisingly synthesize new superheavy nuclei.

\section{\label{sec:conclusion}CONCLUSION}

This paper performs PIC simulations to determine the spatial and angular distribution, the energy and charge, the electric field and the beam parameters of Th ions when the polarized laser is applied to the nm-level thickness CT. Th ions produced by the CTs have slightly higher energy than those produced by the FT, especially for the CT.AL. Th ion beam generated by LPL has remarkably higher energy, charge, beam intensity, beam emittance and energy conversion efficiency compared to that generated by CPL. 
\begin{itemize}
\item A GeV-level, around 0.4 MA and 0.05 mm$\times$mrad Th ion beam can be obtained when the circularly polarized SEL 100 PW laser is applied to the 250 nm either FT or CT.AR.
\item A ten GeV-level, around 1.2 MA and 0.2 mm$\times$mrad Th ion beam can be obtained when the linearly polarized SEL 100 PW laser is applied to the 250 nm the CT.AL.
\item A hundred GeV-level, around 0.6 MA and 6 mm$\times$mrad Th ion beam can be obtained when the linearly polarized SEL 100 PW laser is applied to the 50 nm FT. 
\end{itemize}
The above options could be of assistance to adjust the laser polarization and the shape of metal target to control ion beams according to specific experimental requirements. High-quality Th ion beams is most efficiently achieved by using a mid-thickness CT.AL, which is easy to prepare due to the surface roughness produced by machining can be comparable to the height of the conical target used in this work, while the optimal angularity still requires further detailed research to confirm. These findings provide useful references for the impact of non-uniform metal targets on laser-ion experiments \cite{rubovivc2021measurements, wang2021super}, offer new insights for improving the primary beam quality in laser particle accelerators and are relevant to the approaches for enhancing reaction cross-sections in new superheavy nucleus synthesis using high-quality, high-charge and high-intensity ion bunches by multi-nuclei collisions \cite{wanqing2024multi, niu2021systematics, zhang2024possibilities}.

\begin{acknowledgments}
This work is supported by the National Key Research and Development Program of China (No. 2022YFA1602404, No. 2022YFA1602402), the Strategic Priority Research Program of the CAS under Grant No. XDB34030000, the National Natural Science Foundation of China (No. 12475134, No. 12235003, No. U1832129), the Youth Innovation Promotion Association CAS (No. 2017309), and Natural Science Foundation of Henan Province (Grant No. 242300421048).
\end{acknowledgments}


\end{document}